\pdfoutput=1
\documentclass{JINST}
\usepackage{cite}

\title{Multi-channel front-end board for SiPM readout}

\author{M.~Auger, A.~Ereditato, D.~Goeldi~,  I.~Kreslo\thanks{Corresponding author.}, D.~Lorca, M.~Luethi, C.~Rudolf~von~Rohr, J.~Sinclair, and M.~S.~Weber\\
Laboratory for High Energy Physics\\
Albert Einstein Center for Fundamental Physics\\
Universit\"{a}t Bern, Switzerland\\
E-mail: \email{igor.kreslo@cern.ch}}

\abstract{We describe a novel high-speed front-end electronic board (FEB) for interfacing an array of 32 Silicon Photo-multipliers (SiPM) with a computer. The FEB provides individually adjustable bias on the SiPMs, and performs low-noise analog signal amplification, conditioning and digitization. It provides event timing information accurate to 1.3~ns RMS. The back-end data interface is realized on the basis of 100~Mbps Ethernet. The design allows daisy-chaining of up to 256 units into one network interface, thus enabling compact and efficient readout schemes for multi-channel scintillating detectors, using SiPMs as photo-sensors.}


\begin{document}
\section{Introduction}
Geiger-mode multi-pixel avalanche photo-diodes, also known as Silicon Photo-multipliers (SiPMs) are gradually taking over the market of photon-counting sensors from the vacuum photo-multiplier tubes (PMT). Compared to PMTs, SiPMs are robust, insensitive to vibrations and magnetic field, faster, require much less electric power and provide excellent single-photon resolution. 
Several examples of charged particle detectors utilizing SiPMs to detect scintillating light can be seen here 
\cite{T2K,SBND,DUNE}. 

The front-end electronic board (FEB) design is driven by requirements of the Cosmic-Ray Tracking subsystem of the Short Baseline Near Detector (SBND) Liquid Argon TPC \cite{SBND}.

The subsystem is constructed of 142~panels with strips of scintillator. The panel contains 16 strips, each equipped with 2 wavelength-shifting fibers. The fibers guide converted scintillation light to the end of the strip, where it is read by a pair of SiPMs. The module contains 32~SiPMs to read out. The FEB is designed to serve one such module.
The coincidence between signals from two SiPMs from the same strip allows significant reduction of the dark pulse rate.

The summary of the functionality implemented in the FEB is listed below:
\begin{enumerate}
\item Provides bias voltage in the range of 62-82~V (extendable to 20-90~V) individually adjustable for each of 32 MPPCs;
\item Amplifies and perform shaping of the MPPC output pulse on each of 32 channels;
\item Performs discrimination of shaped signal at configurable level from 0 to 50 SiPM photo-electrons;
\item Provides basic coincidence of signals from each pair of adjacent channels (optional);
\item Allows to trigger only on events that are validated by external signals, such as, by an event in a group of other FEBs;
\item Generates trigger for digitization of the signal amplitude;
\item Generates time stamp w.r.t. the input reference pulse with an accuracy of 1.3~ns;
\item Performs digitization of signal amplitude of each of the 32 channels;
\item Provides on-board data buffering; 
\item Provides efficient back-end communication based on Ethernet standard;
\item Allows firmware upgrade via back end Ethernet link.
\end{enumerate}
The power requirements are: +5V with the consumption ranging from 450~mA to 550~mA depending on channel configuration.

\section{General view and block-scheme}
\begin{figure}[h!]
\centering
\includegraphics[width=\textwidth]{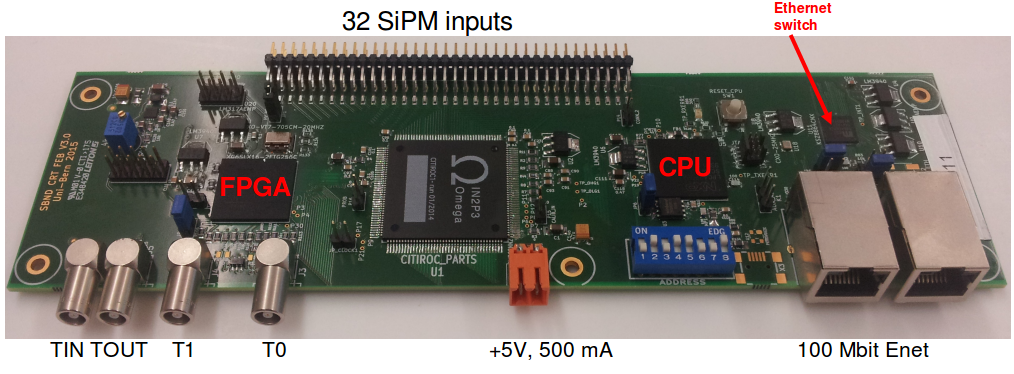}    
\caption{General view of the Front-End Board.}
\label{fig-view}
\end{figure}

The board is realized in a custom compact form-factor suitable for mounting at the edge of a scintillating detector as shown in Figure \ref{fig-view}. At one side a 72-pin SiPM connector is located, the other side hosts a power connector, two RJ45 network jacks and four LEMO connectors for reference and control signals.

\begin{figure}[h!]
\centering
\includegraphics[width=\textwidth]{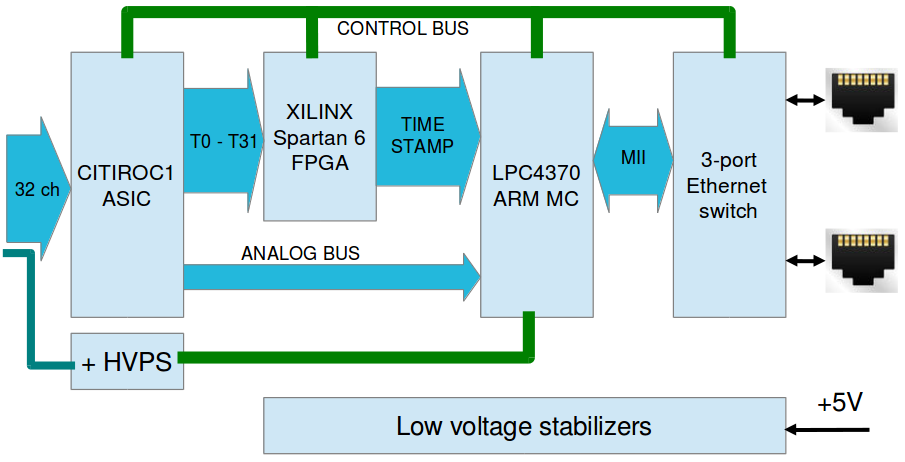}    
\caption{Block-scheme of the Front-End Board.}
 \label{fig-block}
\end{figure}

The block-scheme of the board is shown in Figure \ref{fig-block}.
The analog input signal is processed by a CITIROC 32-channel ASIC from Omega \footnote{\texttt{http://omega.in2p3.fr/}}\cite{citiroc}. For each channel the chip provides charge amplifier with configurable gain, fast shaping with a peaking time of 15~ns and slow shaping with configurable peaking time from 12.5~ns to 87.5~ns. Signals from the fast shaper are discriminated at configurable level and produce digital signals (T0-T31) for triggering an event. These 32 signals are routed to the XILINX Spartan-6 FPGA chip, where the basic input coincidence and event triggering logic is realized. The analog signals for all channels can be stored in the ASIC Sample-and-Hold (S/H) circuit and multiplexed to a single analog output. This output is then routed to the ADC (part of NXP LPC4370 ARM micro-controller chip).

\section{Triggering logic}       

\begin{figure}[h!]
\centering
\includegraphics[width=\textwidth]{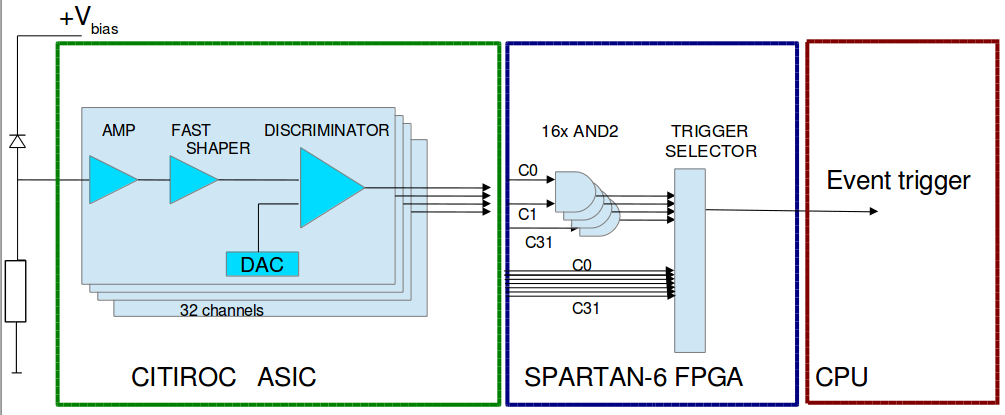}    
\caption{Block-scheme of Front-End Board triggering circuit.}
\label{fig-trig}
\end{figure}

In the Figure \ref{fig-trig} a block-scheme of the trigger formation circuit is shown. Within the CITIROC for each of the 32 channels a charge amplifier with configurable gain and dynamic range from 1 photo-electron (p.e.) to 2000~p.e. (at MPPC gain of 10$^6$) is followed by a fast RC-CR shaper with peaking time of 15~ns. The shaped signal is binarized by a discriminator. The discriminator threshold is supplied by a common 10-bit DAC plus a 4-bits DAC for the fine adjustment of individual channels. Each of these 32 digital trigger signals (C0 to C31) are routed to the FPGA where they are paired with an AND logic to form coincidence signals for each SiPM pair (C0\&C1, C2\&C3 etc.).
The OR of the resulting 16 signals, together with each individual channel trigger signal C0-C31, is fed to the trigger selection logic, which is configurable via back-end interface. Any combination of these signals can be selected to produce a primary event trigger for the CPU interrupt input.
Within the FPGA this signal also triggers the generation of the event time stamp.
Each of 32 channels can be individually enabled or disabled by the CITIROC configuration bit stream.
This bit stream is produced by the on-board CPU on the basis of a configuration command received via Ethernet link from the host computer.

\begin{figure}[h!]
\centering
\includegraphics[width=0.8\textwidth]{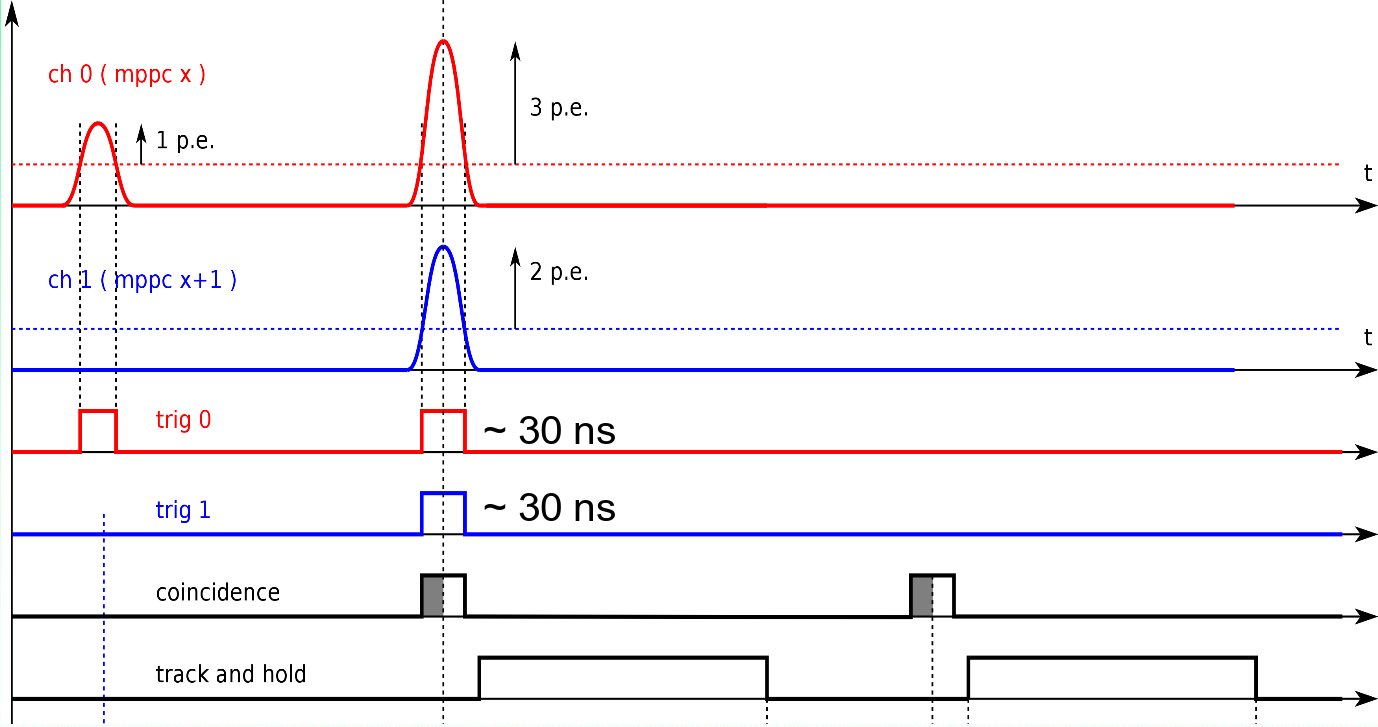}    
\caption{Timing diagram of the triggering circuit. The first event on ch0 (red) has no coincidence with ch1 (blue), so does not produce a trigger. The Second event on ch0 in coincidence with ch1 triggers a readout cycle.}
\label{fig-trig1}
\end{figure}

The timing diagram of the circuit is shown in Figure \ref{fig-trig1}. The time window of generated odd-even channel coincidence varies from 0 to 30~ns depending on the amplitude of the input pulse w.r.t the discriminator threshold. After a delay of 50~ns the primary event trigger sets the HOLD signal to the Sample-and-Hold circuit to memorize instantaneous signal levels at all 32 channels at the moment of the top of the peak (see next Section). The same signal is output at the "TOUT" LEMO connector (Figure \ref{fig-view}). The HOLD signal is kept for at least 150 ns to define the coincidence window.
If during this period the circuit detects a high level at the input "TIN" LEMO connector, the event is considered valid and the HOLD signal is kept high until the CPU finishes the digitization cycle and resets it to initial state.  If no signal is received in "TIN" during 150~ns the HOLD signal is reset by the FPGA and the event is discarded. The "TOUT" signal is reset to zero in any case after 150~ns. This input has an optional on-board weak pull-up resistor, allowing operation without the external signal. Such functionality allows for a hardware coincidence between the event generated by the FEB and other external events.

\section{Bias generator and analog signal readout}       

\begin{figure}[h!]
\centering
\includegraphics[width=\textwidth]{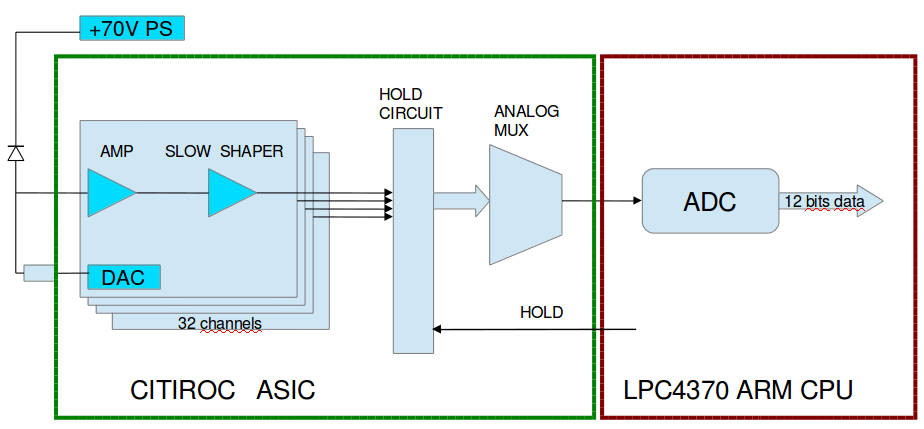}    
\caption{Block-scheme of analog signal processing circuit.}
\label{fig-analog}
\end{figure}

\begin{figure}[h!]
\centering
\includegraphics[width=0.7\textwidth]{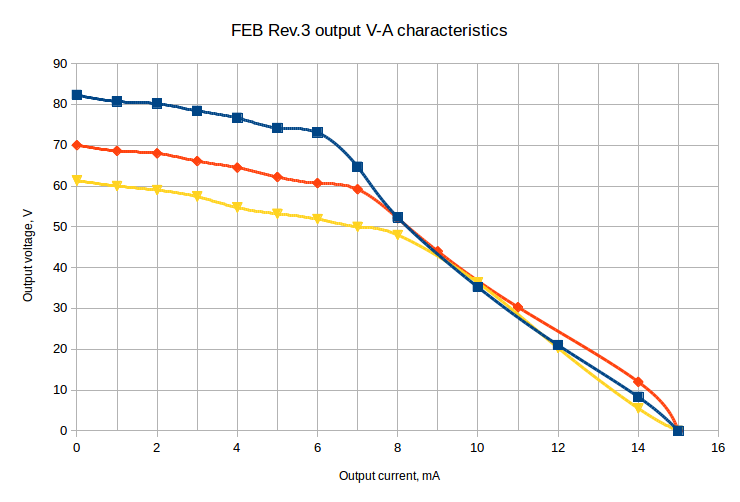}    
\caption{Output Volt-Ampere characteristics of the bias power supply for 62V (yellow), 72V (red) and 82V (blue) bias settings.}
\label{fig-outva}
\end{figure}

The block-scheme of the analog signal processing circuit is shown in Figure \ref{fig-analog}.
The bias voltage is generated by switching a stabilized power supply circuit operating at 100 MHz frequency.
The voltage can be adjusted from 62V to 82V by the trimmer resistor. The range can be extended to 20V-90V by replacing divider resistors in the circuit. This voltage is common for all 32 SiPMs connected to the FEB inputs. The output V-A characteristics of the bias power supply are shown in Figure \ref{fig-outva}. The power supply can be enabled or disabled by a dedicated signal generated by the on-board CPU.  The individual bias voltage adjustment is performed by an 8-bits DACs within the CITIROC. The DAC's positive output levels are supplied to the signal lines as a DC-offset; therefore, increasing this voltage reduces the effective bias for individual SiPMs. The full DAC range is +0.5V to +4.5 V.

\begin{figure}[h!]
\centering
\includegraphics[width=0.8\textwidth]{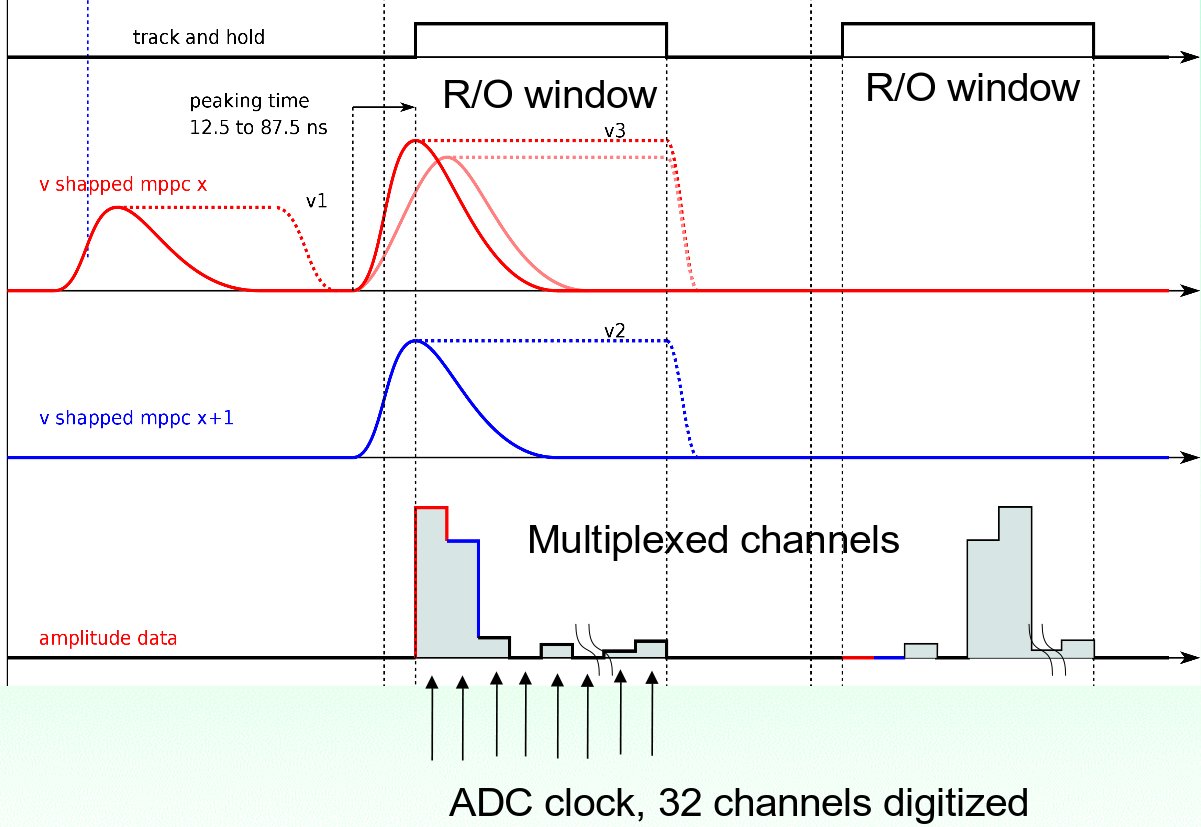}    
\caption{Timing diagram of the analog signal processing circuit. The first event on ch0 (red) has no coincidence with ch1 (blue) and does not produce a trigger. The second event on ch0 in coincidence with ch1 triggers readout cycle. Peak values on all channels are stored by a track-and-hold circuit for a period needed to multiplex them to a common analog output and digitization (R/O window).}
\label{fig-analog1}
\end{figure}

Amplified charge pulses from SiPMs are shaped by a slow RC-CR shaper with configurable peaking time (12.5~ns to 87.5~ns). This time is adjusted in such a way that the HOLD signal, delayed by 50~ns w.r.t. event, has its rise flank at the flat top of the peak of the shaper output pulse, minimizing the noise due to time jitter. The amplitudes at the 32 shaper outputs are latched at S/H circuit and routed to an analog multiplexer. When the CPU receives the trigger interrupt it initiates the readout cycle. The timing diagram is shown in Figure \ref{fig-analog1}. The solid colored lines represent the shaper outputs, the dotted lines the outputs of the S/H circuits. The event illustrated in Figure \ref{fig-analog1} is triggered by the coincidence between two neighboring channels (x and x+1). The CPU controls multiplexing of all 32 outputs via a single line that is routed to the 12-bits ADC input. In Figure \ref{fig-analog1} only 8 channels multiplexing is shown for simplicity.
Once all 32 channels are digitized and stored in the event buffer, the CPU sends the reset signal to the FPGA and completes the readout. Each of the 32 charge ampthe lifiers can be individually enabled or disabled by the CITIROC configuration bit stream.

\begin{figure}[h!]
\centering
\includegraphics[width=0.6\textwidth]{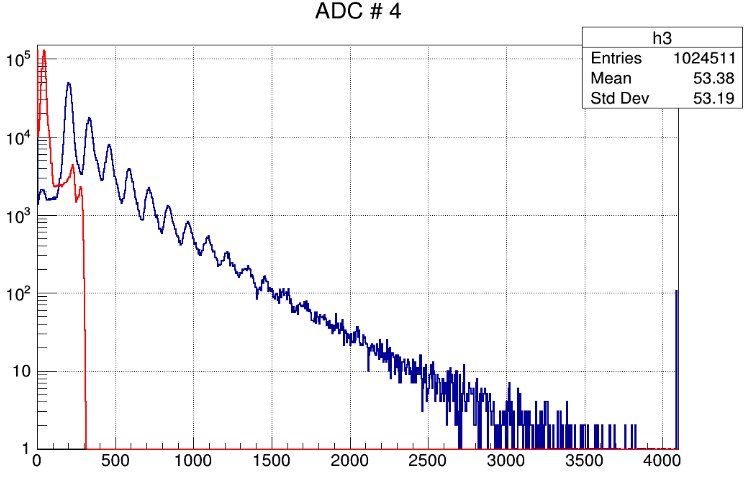}    
\caption{Typical performance of the analog signal processing circuit (with a Hamamatsu S12825-050P MPPC). The amplitude spectrum for dark counts with the discrimination threshold at 0.5 p.e. is shown in blue.
The red spectrum shows the location of the pedestal (it comes from the non-triggering channel).}
\label{fig-analog2}
\end{figure}

The performance of the analog readout circuit is illustrated in Figure \ref{fig-analog2}, where the amplitude spectrum for dark counts with a discrimination threshold at 0.5~p.e. is shown in blue.
The red spectrum shows the location of the pedestal when triggered by some other channel.

\section{Time stamp generator}       
The Time-to-Digit Converter (TDC) of the time stamp generator is composed of the coarse counter, working at the clock frequency of 250 MHz, and the delay-chain interpolator improving accuracy down to 1~ns. 
The solution is based on the approach published in \cite{tdc1}. 

\begin{figure}[h!]
\centering
\includegraphics[width=\textwidth]{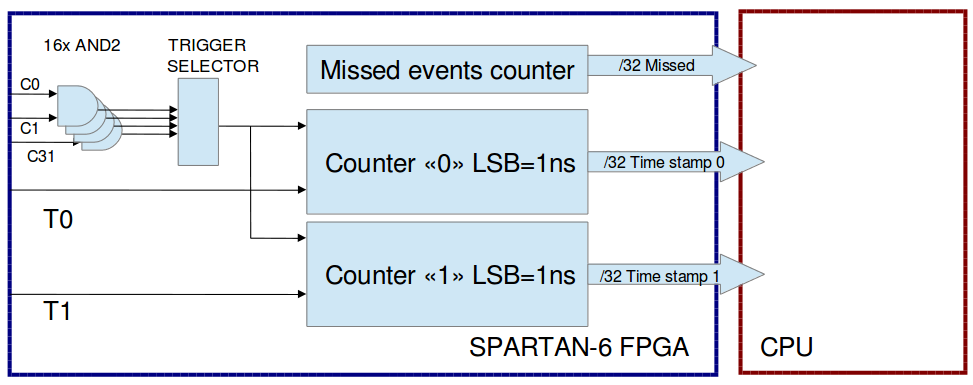}    
\caption{Block-scheme of time stamp generation circuit.}
\label{fig-tdc}
\end{figure}

\begin{figure}[h!]
\centering
\includegraphics[width=0.6\textwidth]{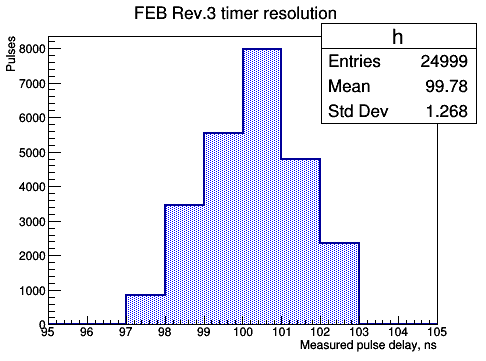}  
\caption{Accuracy of the time-stamp generator. Events are delayed by 100 ns w.r.t. the reference signal.}
\label{fig-tdc2}
\end{figure}

For each event the FEB is capable of recording two independent time stamps w.r.t the positive flank on "T0" and "T1" LEMO inputs (Figure \ref{fig-tdc}). Each time stamp is a 32-bits word, having time information in 30 Least Significant Bits (LSB) represented in Gray code. The resolution of the circuit is illustrated in Figure \ref{fig-tdc2}, where the events arrive 100~ns after the T0 reference signal. 

Two Most Significant Bits (MSB) are used for flagging special events. Two special events are foreseen. The first is the arrival of the reference signal at either "T0" or "T1" inputs (time reference event). For such an event the time passed since the previous time reference event is recorded, and the timing circuit is reset to zero. This allows to measure the period between reference signals and, in case the real period is highly stable and accurate, the measured period allows to derive a deviation of the internal on-board oscillator frequency from its nominal value. If this deviation is known it can be applied offline to scale all time stamps between two reference pulses to recover accuracy.

The second special event happens in the absence of the reference pulse for more than 1074~ms, which leads to overflow of the coarse counter. This situation is flagged and can be used by offline software to invalidate the time stamp until the presence of the reference pulse is restored.

\begin{figure}[h!]
\centering
\includegraphics[width=\textwidth]{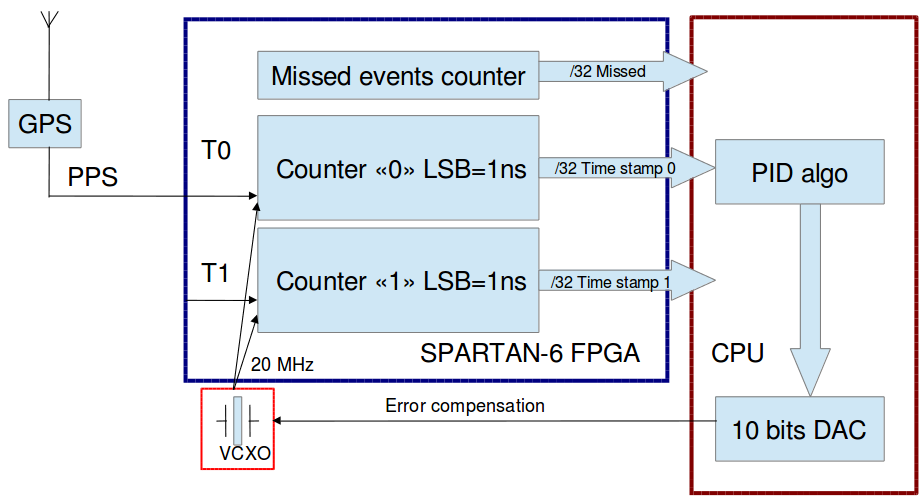}    
\caption{Block-scheme of the oscillator control loop of the time stamp generation circuit.}
\label{fig-tdcloop}
\end{figure}

\begin{figure}[h!]
\centering
\includegraphics[width=0.95\textwidth]{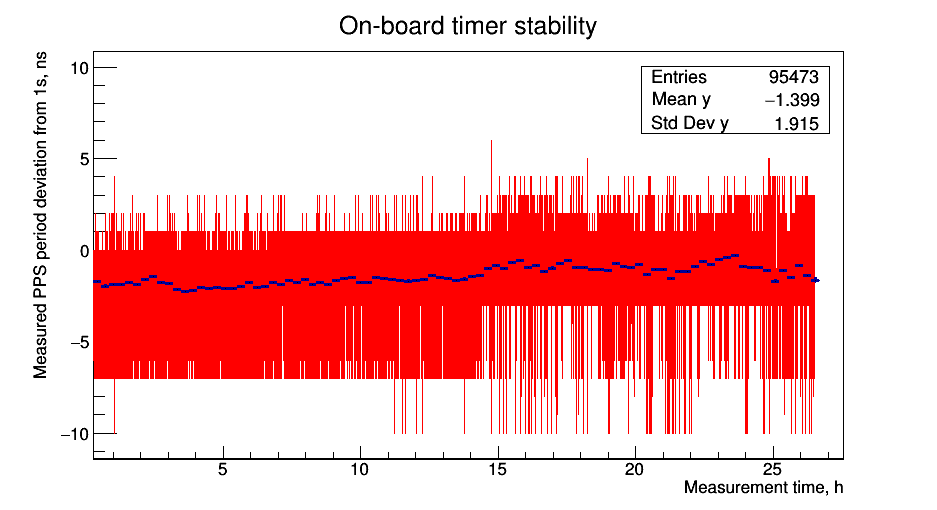}  
\caption{Stability of the time-stamp generator disciplined oscillator.}
\label{fig-tdc1}
\end{figure}

The 20~MHz temperature-compensated voltage-controlled crystal oscillator (VCXO) is used as a source of the reference clock for the FPGA and timing circuit. The feedback voltage for the VCXO is generated by a 10-bits DAC under control of the on-board CPU (Figure \ref{fig-tdcloop}). The frequency correction signal is derived on-board, based of measured periods between reference pulses supplied to the "T0" input from the high-stability GPS-disciplined pulse-per-second (PPS) generator. The performance of the timer and oscillator control loop can be seen at Figure \ref{fig-tdc1} where the measured period between PPS reference pulses is shown. The PPS pulse for this measurement is supplied by an external high-stability oscillator disciplined with the PPS from U-Blox M8T\footnote{\texttt{http://www.u-blox.com}} high-accuracy GPS receiver for timing applications. The correction signal is derived after each 20 successive measurements of the PPS period. After about a minute the control loop is stabilized and maintains stability of the on-board oscillator within $1.9\times10^{-9}$ 1-s Alan deviation.

Together with two 32-bits words from the two time stamp generators, an additional word is present in time stamp data. This word contains the number of extra primary event triggers that occurred during the readout cycle of one event. These events will be missed by the FEB and knowledge of their number allows to measure the current acquisition inefficiency.

Once the time stamp for the event is latched in the internal FPGA register the CPU initiates the transmission of this data via a Serial Peripheral Interface (SPI).

\section{Event buffer and back-end Ethernet interface}       

Once the CPU completes digitization of all 32 analog channels and transmission of the time stamp from the FPGA it combines this data into an event and stores it in the internal ring buffer, which has a capacity of 1024 events. The on-board micro-controller contains three CPU cores, each working at 160 MHz clock frequency. One core is taking care of filling the ring event buffer, while the second one is emptying it sending data out via an on-board Ethernet switch when it is requested to do so by the host PC.
If the incoming data rate exceeds the capacity of the back-end interface, the events are overwritten in the ring buffer. This situation is detected by the CPU and the number of overwritten (and therefore lost) events is stored, to be transmitted to the host PC. This number together with the "Missed event" counter in the FPGA contributes to the measured acquisition inefficiency of the FEB.

\begin{figure}[h!]
\centering
\includegraphics[width=0.8\textwidth]{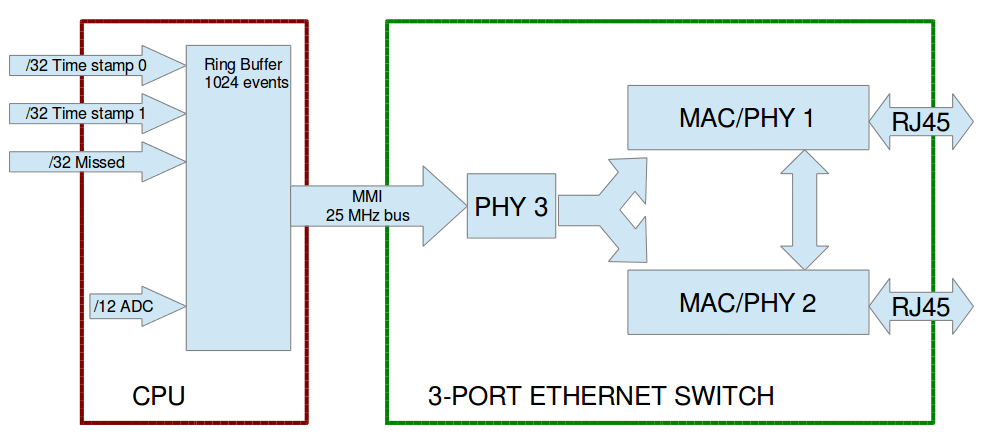}    
\caption{Block-scheme of the back-end data transmission and control interface.}
\label{fig-backend}
\end{figure}

The second CPU core communicates with the host computer via three-port Ethernet switch IC. The switch to the CPU interface (port 3) is a 25 MHz 4-bits MII bus with throughput of 100 Mbps. The other two ports (1 and 2) of the switch have Physical Layer (PHY) transmitters and are connected to two RJ45 Ethernet jacks. The switch is configured to forward ports 1 and 2 in both directions, to accept control commands on these ports and to forward these commands via the port 3 to the on-board CPU (Figure \ref{fig-backend}.
The event data from the CPU is forwarded only to that port from which the FEB received a data transmission request.

Both PHY ports of the switch have MDI/MDIX Auto Cross functionality. Therefore, the user does not need to care about the structure of the connecting Ethernet cables (straight or cross-wired). The switch detects the type of the cable on connection and sets its configuration accordingly.

\begin{figure}[h!]
\centering
\includegraphics[width=0.50\linewidth]{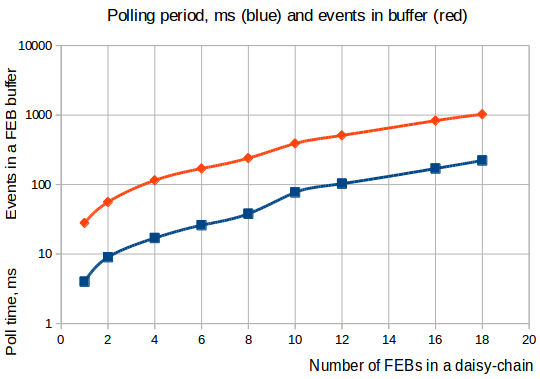}     
\includegraphics[width=0.48\linewidth]{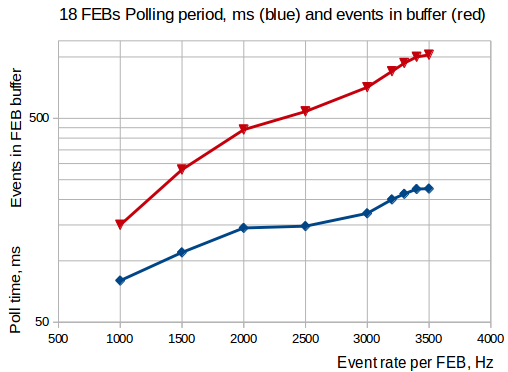}     
\caption{Measured polling period, in ms, and buffer use as a function of the number of FEBs in the chain for a fixed event rate of 3.5 kHz (left) and as a function of the event rate for 18 FEBs in a chain (right).}
\label{fig-flux}
\end{figure}

A test was conducted with several FEBs daisy-chained to one host PC running development data acquisition code. Events were simulated by sending reference pulses to the "T1" trigger input at a variable rate. As the event rate increased, the time to transfer data from all FEBs (polling period) also increased. The maximum number of events accumulated in each FEB internal buffer increases accordingly. This number is limited by the buffer capacity (1024 events). Variations of these two parameters as a function of the number of FEBs in a chain at 3.5 kHz event rate is shown on the left of Figure \ref{fig-flux}. In the right plot of  Figure \ref{fig-flux} the dependence on the event rate with a fixed number of FEBs (18) in the chain is shown.

\begin{figure}[h!]
\centering
\includegraphics[width=0.6\textwidth]{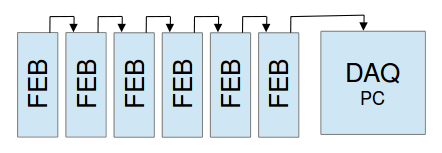}    
\caption{Scheme of daisy-chain communication between several FEBs and host PC.}
\label{fig-chain}
\end{figure}

The MAC address of each board is set by firmware to 00:60:37:12:34:XX where last XX byte can be set from the hardware 8-bits switch array. 
In Figure \ref{fig-chain} an example communication scheme between several FEBs and a host PC is shown. FEBs are daisy-chained and connected to a single Ethernet port of the host PC.

\section{Conclusions}
A novel high-speed front-end electronic board (FEB) for interfacing an array of 32 Silicon Photo-multipliers (SiPM) to a computer has been developed. The FEB incorporates analog and digital processing circuits, provides bias supply, 
signal processing and digitization, buffering and communication with the host computer via a 100 Mbps Ethernet link. The FEB provides event time stamping with 1.3 ns accuracy, w.r.t. an external reference signal, such as a GPS Pulse-Per-Second. The design provides capability of daisy-chaining of up to 256 boards into one network interface cable. The FEB design enables low-power, compact and efficient readout schemes for multi-channel instrumentation with SiPMs as photo-sensors. The FEB is commercialized and available for purchasing from CAEN\footnote{CAEN - Costruzioni Apparecchiature Elettroniche Nucleari S.p.A} (A1702\footnote{http://www.caen.it/servlet/checkCaenDocumentFile?Id=11427}).
\section{Acknowledgments}
We acknowledge financial support of the Swiss National Science Foundation.


\end{document}